# Joint Range-modulator and Spot Optimization for Bragg-peak Proton FLASH Radiotherapy

Jiayue Han[1], Ya-Nan Zhu[1*], Aoxiang Wang[1,2], Wangyao Li[1], Yuting Lin[1], Hao Gao[1*]

[1]Department of Radiation Oncology, University of Kansas Medical Center, USA

[2]Department of Biomedical Engineering, Huazhong University of Science and Technology, Wuhan, China

*Email: yananzhu6@gmail.com, hgao2@kumc.edu

**Acknowledgment**

The authors are very thankful to the valuable comments from reviewers. This research is partially supported by the NIH grants No. R37CA250921, R01CA261964, and a KUCC physicist-scientist recruiting grant.

**Abstract.**

**Background:** Ultra-high-dose-rate (UHDR) radiation therapy has demonstrated promising potential in reducing toxicity to organs-at-risk (OARs). Proton therapy is uniquely positioned to deliver UHDR by leveraging the Bragg peak in conjunction with patient-specific range modulators (PSRMs) to generate a spread-out Bragg peak (SOBP). Existing proton FLASH (pFLASH) treatment planning workflows typically follow a two-step process: (1) generating a multi-energy intensity-modulated proton therapy (IMPT) plan to determine spot weights, and (2) subsequently converting this plan into a single-energy pFLASH delivery using PSRM optimization. However, the intrinsic coupling between spot weight distribution and PSRM design has not been fully investigated, which may limit the achievable dosimetric and radiobiological advantages of pFLASH therapy.

**Purpose:** This work proposes a novel alternating optimization framework—Joint Range-Modulator and Spot Optimization (JRSO)—that simultaneously optimizes the PSRM and spot weights to improve the plan quality of conformal pFLASH therapy.

**Methods:** Compared to the conventional method, JRSO does not require a one-to-one correspondence between beam spots and PSRM pins. Proton beam diffusion from the delivery system to the PSRM is modeled by Gaussian function. To achieve better plan quality, starting from an initial solution derived from a conventional IMPT plan, JRSO iteratively updates the PSRM design and spot weights in an alternating manner. This process progressively refines both parameters while ensuring compliance with practical delivery constraints, such as the minimum monitor-unit (MMU) requirement.

**Results:** JRSO obtained improved plan quality compared to the conventional method. For example, in a head-and-neck (HN) case, JRSO reduced the objective function value from 0.46 to 0.26, lowered the maximum target dose from 117.6% to 107.1%, improved the conformity index from 0.74 to 0.87, and decreased the region-of-interest (ROI) effective dose from 6.50 Gy to 6.10 Gy.

**Conclusion:** A new optimization method JRSO is proposed for conformal pFLASH radiotherapy. It outperforms the conventional approach and may extend the applicability of PSRM to more complex clinical scenarios, particularly those involving misalignments between beam spots and pins. Numerical results demonstrate the robustness and efficiency of the new method.

1. **Introduction**

FLASH radiotherapy (RT) delivers ultra-high dose of radiation within an extremely short time frame, effectively damaging tumor cells while reducing normal tissue toxicity by approximately 20%-40% compared to conventional dose rates [1-4]. Although the biological mechanisms underlying the FLASH effect remain unclear [5-8], initial studies suggest that dose rates exceeding 40 Gy/s and dose exceeding 8 Gy [3] can induce this effect. Subsequent research has indicated a broader minimum dose rate range of 10–100 Gy/s [4,9-11]. FLASH beams can be generated using electrons [12], X-rays [13], or protons [14], with protons being particularly promising for clinical translation due to their ability to penetrate deep-seated tumors while reducing the entrance dose.

However, at the FLASH dose rate, conventional intensity-modulated proton therapy

(IMPT) plans cannot be delivered using current pencil beam scanning (PBS) systems. As is discussed in [15], changing the beam energy layer takes approximately one second, which can adversely affect the dose rate. To address this challenge, patient-specific range modulators (PSRM) have been designed to modulate spot weight [16-17]. PSRM are an innovative component in conformal proton FLASH (pFLASH) therapy, designed to enhance the precision and conformity of dose delivery. By tailoring the modulation of proton beams to the individual geometry and depth of a patient's tumor, PSRMs enable the formation of conformal dose distributions while preserving the ultra-high dose rates required to achieve the FLASH effect.

Current pFLASH planning workflows typically begin with the optimization of a multi-energy IMPT plan to determine reference spot weights, followed by conversion into a single-energy pFLASH plan using a patient-specific range modulator (PSRM) [17–20]. However, these approaches often overlook the intrinsic coupling between spot weight distribution and PSRM geometry. To the best of our knowledge, [15] represents the first attempt to optimize the PSRM and spot weights separately. Nonetheless, their method assumes a one-to-one correspondence between beam spots and the pins of the range modulator—a simplification that may result in misalignments when the number of spots does not match the number of pins.

To address these challenges, we propose a novel method termed Joint Range-modulator and Spot Optimization (JRSO). JRSO is an integrated optimization framework that simultaneously optimizes the shape of the range modulator and the intensity of the spots. Unlike previous approaches, it is specifically designed to account for potential misalignments between beam spots and modulator pins, thereby improving the robustness and accuracy of conformal pFLASH treatment planning.

## 2. Methods and Materials

*2.1. Problem formulation*

The basic objective for treatment planning in radiation therapy is to deliver sufficient dose to the target region while minimizing the dose exposure to the OARs. A typical IMPT plan solves the optimization problem as follows:

$$\min_{z \in \mathbb{R}^N} f(z) = \sum_{s=1}^{S} \omega_s \left\| D_{\Omega_s(z)} z - d_{\Omega_s(z)} \right\|^2, \qquad (1)$$

$$s.t. \ z_i \in \{0\} \cup [g, +\infty)$$

where $d_{\Omega_s(z)}$ is the target dose distribution, $z$ is the spot weight to be optimized and $D_{\Omega_s(z)}$ is dose influence matrix for the *s-th* structure. The function $f(z)$ consists of three parts: (1) The least square terms of the targets and the OARs, which measures the distance between the optimized dose distribution and the target dose distribution; (2) Dose volume histogram (DVH) objective: e.g., DVH-max-dose constraints $D_{p\%} \leq c$ (for OAR), which implies that no more than $p\%$ of the OAR volume receive dose more than $c$; (3) DVH-min-dose constraints $D_{p\%} \geq c$ (for target), which implies that no less than $p\%$ of the target volume receives a dose equal at least $c$. Here for simplicity, we write them in the same formulation. The constraint in equation (1) is referred to as the Minimum Monitor-Unit (MMU) constraint, which states that the beam intensity is nonnegative and larger than $g$ when it is positive [21].

The principles underlying the overall framework is illustrated in Figure 1 (a). In the proposed setting, the single energy FLASH beam passes through a PSRM to modulate the dose distribution inside the target. An example for 3D model of the PSRM is presented in Figure 1 (c). The PSRM comprises multiple small pins, each shaped like a pyramid, as shown in Figure 1 (b). The ability of a PSRM pin to attenuate proton energy increases with its height. In this study, a discrete set of candidate pin heights is predefined, and the corresponding area associated with each height is treated as an optimization variable. This formulation enables flexible modulation of the proton range while maintaining a physically realizable modulator design.

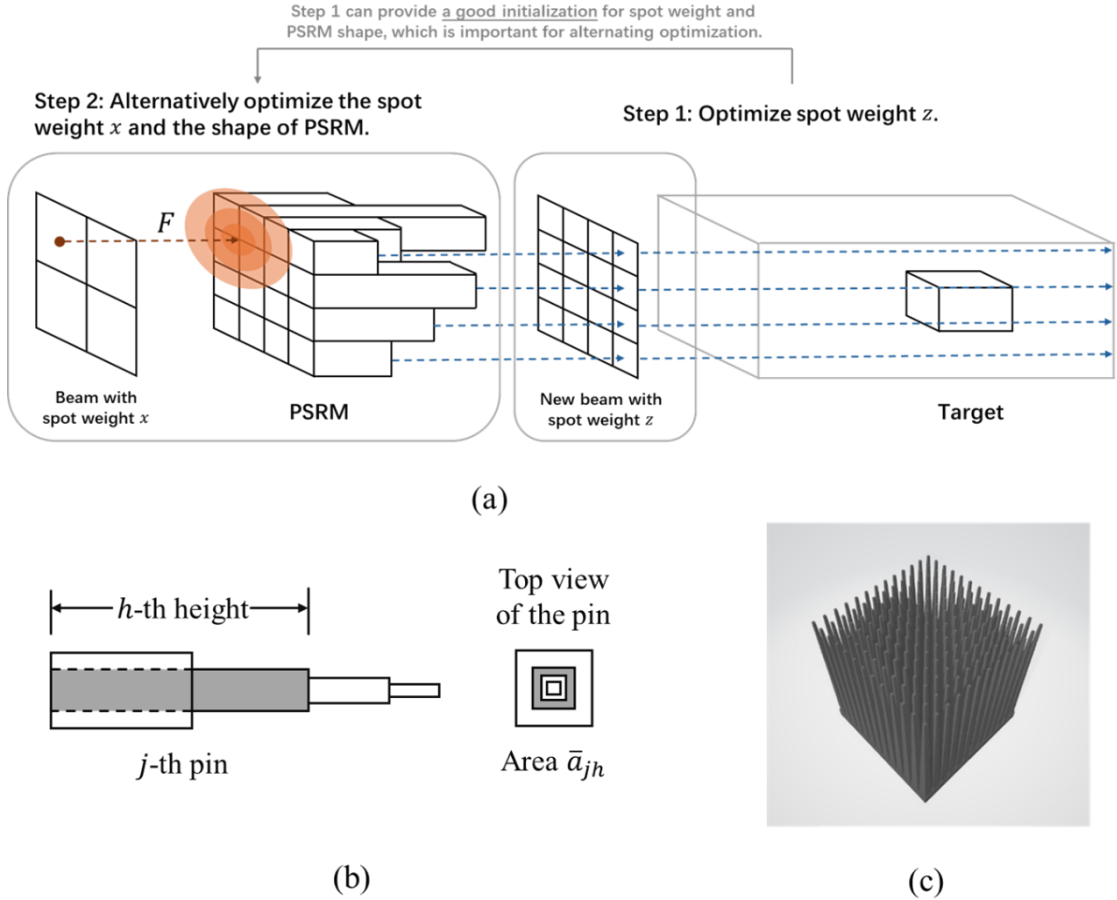

**Figure 1.** Illustration of the JRSO framework, PSRM design, and experimental validation. (a) Diagram for the problem formulation and the JRSO. (b) Diagram of one single pin. (c) 3D plot of a PSRM.

*2.2 JRSO*

In this section, we give the mathematical model of JRSO. Define $x \in \mathbb{R}^I$ as the spot weight for the beam coming out of the PBS, as shown in Figure 1. Let $y \in \mathbb{R}^J$ be the actual spot weight that goes into each pin of the PSRM. Note that the number of spots and the number of pins may not be the same, which suggests that the dimension of $x$ and the dimension of $y$ may not be equal. To account for beam scattering in air and address potential misalignments between the beams and each pin of PSRM, a coefficient matrix $F \in \mathbb{R}^{J \times I}$ is introduced. Let $i \in [I], j \in [J]$, here $[I]$ denotes the set of indices from 1 to $I$, i.e., $[I] = \{1, 2, \ldots, I\}$. The element in the $j$-th row and $i$-th column of $F$ represents the proportion of protons originating from the $i$-th beam that subsequently enter the $j$-th pin. This formulation accommodates the one-to-many correspondence between beams and pins and accounts for possible misalignments of beam centers.

Adopting the gaussian model, the weight matrix can be defined as the follows: consider the coordinate of $i$-th spot and $j$-th pin $P_{xi} \in \mathbb{R}^2, P_{yj} = (P_{yj}^1, P_{yj}^2) \in \mathbb{R}^2$, the entry $F_{ij}$ of matrix $F$ is defined by

$$F_{ji} = \iint \frac{1}{2\pi |\Sigma|^{\frac{1}{2}}} \exp\left[-\frac{1}{2}(P_{xi} - P_{yj})^T \Sigma^{-1}(P_{xi} - P_{yj})\right] dP_{yj}^1 dP_{yj}^2, \tag{2}$$

where $\Sigma = \begin{pmatrix} 3 & 0 \\ 0 & 3 \end{pmatrix}$ is selected based on the lateral profile, and the integral is computed on the base area of $j$-th pin. With the definition of $F$, $y$ is given by

$$y = Fx.$$

The area corresponding to each height for each pin is represented by a matrix $\bar{a} \in \mathbb{R}^{J \times H}$, where $H$ is the number of candidate heights. Element $\bar{a}_{jh}$ represents the area of the $j$-th pin at height $h$, as shown in Figure 2 (a). To avoid the abuse of notations, we also denote $a \in \mathbb{R}^N, N = J \times H$ as a substitute of $\bar{a}$ when it is reshaped to a column vector. By adjusting the area corresponding to each height, it is possible to control the effective beam intensity that reaches the target. To achieve an optimal treatment plan, the spot weight $x$ and the pin area $a$ must be optimized simultaneously.

With the definition of $x$ and $a$, the mathematical model takes the following form

$$\min_{x,a} G(x, a) = \sum_{s=1}^{S} \omega_s ||D_{\Omega_s(x,a)}[(I_H F x) \circ a] - d_{\Omega_s(x,a)}||^2, \quad (3)$$

$$s.t. \begin{cases} x_i \in \{0\} \cup [g, +\infty) \\ \sum_h \bar{a}_{jh} = 1 \\ \bar{a}_{jh} \in [0, +\infty) \end{cases}$$

where $i \in [I], j \in [J], h \in [H]$, $\circ$ refers to the Hadamard product (i.e. entry wise product) and $I_H$ is defined as

$$I_H = \begin{bmatrix} \mathbb{I}^{J \times J} \\ \vdots \\ \mathbb{I}^{J \times J} \end{bmatrix} \in \mathbb{R}^{N \times J}.$$

Here $\mathbb{I}^{J \times J}$ refers to a $J$-by-$J$ identity matrix and $N = J \times H$.

*2.3. Optimization algorithm*

The objective in (3) is nonlinear and the two variable is coupled. We propose to solve it by the following alternating minimization strategy.

---
**Algorithm 1**: JRSO
---
**Input**: Parameters of the problem.
Initialize $x^0, a^0$.
**For** $k = 1, 2, \ldots, K$
  Solve $x^k = \underset{x}{\text{argmin}}\, G(x, a^{k-1})$.
  Solve $a^k = \underset{a}{\text{argmin}}\, G(x^k, a)$.
**End**
**Output** $x^K, a^K$.

---

In the following sections, we elaborate on each subproblem. Specifically, the optimization of beam intensity $x$ is presented in section 2.3.1, the optimization of the area $a$ of each pin in the range modulator is detailed in section 2.3.2.

2.3.1 Optimization for beam intensity

With basic matrix transformation, the subproblem for $x^k$ can be rewritten as

$$x^k = \underset{x}{\text{argmin}} \sum_{s=1}^{S} \omega_s ||D_{\Omega_s(x,a^{k-1})}[(I_H F x) \circ a^{k-1}] - d_{\Omega_s(x,a^{k-1})}||^2, \quad (5)$$

s.t. $x_i \in \{0\} \cup [g, +\infty)$.

For simplicity, we rewrite the objective function in (5) concisely as follows:

$$x^k = \underset{x}{\text{argmin}} \frac{1}{2} \left\| A^k \left( \Omega(x, a^{k-1}) \right) x - b \left( \Omega(x, a^{k-1}) \right) \right\|^2, \quad (6)$$

s.t. $x_i \in \{0\} \cup [g, +\infty)$.

and $\Omega(x, a^{k-1})$ is the active index selected based on $x$ and $a^{k-1}$.

Equation (6) can be reformulated and solved using the alternating direction method of multipliers (ADMM) [22-28]. The detailed updating rule is given in Algorithm 2 and the derivation of the ADMM formulation is presented in the Appendix.

---

**Algorithm 2**: ADMM for optimizing $x$

**Input**: $a^{k-1}, x^{k-1}, \mu^0, \lambda^0$, MMU threshold $g$, proper $\rho > 0$.

Let $\hat{x}^0 = x^{k-1}$.

**For** $m = 1, 2, \dots, M$

  Update $\Omega(\hat{x}^{m-1}, a^{k-1})$

  $\hat{x}^m = \left( A^k \left( \Omega(\hat{x}^{m-1}, a^{k-1}) \right)^T A^k \left( \Omega(\hat{x}^{m-1}, a^{k-1}) \right) + \rho I \right)^{-1}$
  $\left( A^k \left( \Omega(\hat{x}^{m-1}, a^{k-1}) \right)^T b \left( \Omega(\hat{x}^{m-1}, a^{k-1}) \right) + \rho(\mu^{m-1} - \lambda^{m-1}) \right).$

  $\mu^m = \begin{cases} \max(\hat{x}^m + \lambda^{m-1}, g), & \hat{x}^m + \lambda^{m-1} \geq \frac{g}{2}, \\ 0, & \text{otherwise.} \end{cases}$

  $\lambda^m = \lambda^{m-1} + \hat{x}^m - \mu^m$.

**End**

Let $x^k = \hat{x}^M$.

**Output** $x^k$.

---

*2.3.2 Optimization for range modulator*

Similar as above, after conducting matrix transformation, the subproblem for $a^k$ takes the following form

$$a^k = \underset{a}{\text{argmin}} \sum_{s=1}^{S} \omega_s \left\| D_{\Omega_s(x^k, a)} \left[ (I_H F x^k) \circ a \right] - d_{\Omega_s(x^k, a)} \right\|^2, \quad (7)$$

$$\text{s.t.} \begin{cases} \sum_h \bar{a}_{jh} = 1 \\ \bar{a}_{jh} \in [0, +\infty). \end{cases}$$

By introducing matrix $C$

$$C = \begin{bmatrix} 1 & \cdots & 1 & 0 & \cdots & 0 & \cdots & 0 & \cdots & 0 \\ 0 & \cdots & 0 & 1 & \cdots & 1 & \cdots & 0 & \cdots & 0 \\ \vdots & & \vdots & \vdots & & \vdots & & \vdots & & \vdots \\ 0 & \cdots & 0 & 0 & \cdots & 0 & \cdots & 1 & \cdots & 1 \end{bmatrix}_{J \times N}$$

and analogous to (6), (7) can be rewritten as

$$a^k = \underset{a}{\operatorname{argmin}} \frac{1}{2} \left\| A^k \left( \Omega(x^k, a) \right) a - b \left( \Omega(x^k, a) \right) \right\|^2, \tag{8}$$

$$\text{s.t.} \begin{cases} Ca = \mathbb{1} \\ \bar{a}_{jh} \in [0, +\infty). \end{cases}$$

where $\Omega(x^k, a)$ is the active index selected based on $x^k$ and $a$, $\mathbb{1}$ is an all-one vector.

The ADMM algorithm for solving (8) update as the follows

---

**Algorithm 3**: ADMM for optimizing $a$

**Input**: $x^k, a^{k-1}, \mu^0, \lambda_1^0, \lambda_2^0$, proper $\rho > 0$.

Let $\hat{a}^0 = a^{k-1}$.

**For** $m = 1, 2, \ldots, M$

  Update $\Omega(x^k, \hat{a}^{m-1})$.

  $\hat{a}^m = \left( A^k(\Omega(x^k, \hat{a}^{m-1}))^T A^k(\Omega(x^k, \hat{a}^{m-1})) + \rho C^T C + \rho I \right)^{-1}$
    $\left( A^k(\Omega(x^k, \hat{a}^{m-1}))^T b(\Omega(x^k, \hat{a}^{m-1})) + \rho (C^T \mathbb{1} + C^T \lambda_1^{m-1} + \mu^{m-1} - \lambda_2^{m-1}) \right),$

  $\mu^m = \begin{cases} \hat{a}^m + \lambda_2^{m-1}, & \hat{a}^m + \lambda_2^{m-1} \geq 0, \\ 0, & \text{otherwise}. \end{cases}$

  $\lambda_1^m = \lambda_1^{m-1} + C\hat{a}^m - \mathbb{1}.$

  $\lambda_2^m = \lambda_2^{m-1} + \hat{a}^m - \mu^m.$

**End**

Let $a^k = \hat{a}^M$.

**Output** $a^k$.

---

*2.5 Materials*

The proposed JRSO method was compared to a conventional pFLASH method across three clinical cases: prostate, lung, and head-and-neck (HN). In the conventional method, a reference IMPT plan was initially generated followed by generating the corresponding PSRM and spot weights accordingly, as is described in Algorithm 4 in Appendix. The dose influence matrix for all cases were computed using matRad [29], with 5 mm spot width (in both directions), on 3 mm³ dose grid.

Clinical verified angles are used for the three cases: there are (0°, 180°) for prostate, (0°,

120°, 240°) for lung, and (45°, 135°, 225°, 315°) for the HN. PSRMs were designed accordingly for each beam configuration. The prescribed dose for all the cases is set to be 10 Gy.

To evaluate the performance of the JRSO method under conditions of spot-to-pin misalignment, three distinct spot number configurations were tested for each case. The pin number for each PSRM and the spot number for each beam are shown in Table 1. For each problem setting, the algorithm is run until the maximum iteration number is achieved.

**Table 1.** The setting of pin number and spot number for prostate case, HN case and lung case.

|  | Pin number | Spot number |
|---|---|---|
| Prostate | [143,143] | (1) [116, 116] (total 232)<br>(2) [77, 77] (total 154)<br>(3) [49, 49] (total 98) |
| HN | [103,103,103,103] | (1) [43, 44, 44, 43] (total 174)<br>(2) [22, 24, 24, 23] (total 93)<br>(3) [18, 20, 19, 20] (total 77) |
| Lung | [294, 313, 355] | (1) [134, 140, 162] (total 436)<br>(2) [67, 72, 83] (total 222)<br>(3) [53, 57, 64] (total 174) |

Plan quality was assessed using the conformity index (CI), defined as: $CI = \frac{V_p^2}{\hat{V}_p \times V_{CTV}}$, where $V_p$ represents the volume of the CTV receiving a dose greater than or equal to the prescription dose, $\hat{V}_p$ denotes the total volume receiving a dose greater than the prescription dose, and $V_{CTV}$ is the volume of the CTV. Clinically relevant dose volume histogram (DVH) constraints were employed, and all treatment plans were normalized to ensure $D_{95\%} = 100\%$. The FLASH coverage and the Region of Interest (ROI) effective dose are also evaluated in this study. The ROI, a.k.a CTV1cm is defined as the annulus region of CTV with 1cm extension. FLASH coverage is quantified as the percentage of the ROI receiving a dose greater than 8 Gy while maintaining a dose rate exceeding 40 Gy/s. The ROI effective dose is a biologically weighted dose metric that accounts for the normal tissue-sparing effect of FLASH radiation. Specifically, it is defined as: $D_{ROI} = \frac{1}{N_D}(\|D_f\|_1 \times 0.7 + \|D_{nf}\|_1)$, where $N_D$ is the total number of voxels, $D_f$ represents the dose vector of the voxels that triggers FLASH effect, and $D_{nf}$ is the dose vector of voxels not triggering the FLASH effect. $\|\cdot\|_1$ is L$_1$ norm that is defined as the sum of the absolute values of the entries in a given vector.

## 3. Results

### 3.1. Prostate

The comparison between the conventional method and the proposed JRSO for prostate cancer treatment planning is presented in Table 2. Specifically, in the case with 154 spots, JRSO reduced the objective function value from 0.143 to 0.131, decreased the maximum target dose from 120.6% to 109.6%, and improved the CI from 0.79 to 0.82. Additionally, FLASH coverage increased from 93.4% to 93.5%, while the ROI effective dose was reduced from 6.33 to 6.30. In the case with 122 spots, JRSO further demonstrated its superiority by lowering the objective function value from 0.162 to 0.133, decreasing the maximum dose from 132.8% to 109.7%, and enhancing the CI from 0.80 to 0.83. The FLASH coverage improved from 94.2% to 94.6%, and the ROI effective dose was reduced from 6.40 to 6.38. The most significant improvements were observed in the case with 98 spots, where JRSO reduced the objective function value from 0.259 to 0.231, lowered the maximum dose from 121.8% to 117.3%, and increased the CI from 0.68 to 0.73. FLASH coverage improved from 91.4% to 91.6%, while the ROI effective dose decreased from 6.41 to 6.32. Furthermore, the DVH plots and dose distribution plots in Figure 2 further validate the results. Notably, the performance advantage of JRSO becomes more pronounced as the number of spots decreases, underscoring its robustness and effectiveness in handling challenging conditions with reduced spot counts.

*3.2. Lung*

The comparison between the conventional method and the proposed JRSO method for the lung cancer case is presented in Table 3. For the scenario with 436 spots, JRSO reduced the objective function value from 0.249 to 0.218, decreased the maximum dose from 116.0% to 110.1%, and improved the CI from 0.86 to 0.89. Although the target coverage was 0.6% lower than that achieved by the conventional method, JRSO led to a reduction in the ROI effective dose from 6.33 to 6.19. In the case with 222 spots, JRSO further demonstrated its effectiveness, reducing the objective function value from 0.264 to 0.227, lowering the maximum dose from 114.7% to 111.4%, and enhancing the CI from 0.83 to 0.88. Additionally, JRSO achieved a FLASH coverage of 92.8% and an ROI effective dose of 6.18, both of which surpassed the corresponding values obtained with the conventional method (91.0% and 6.22, respectively). In the most challenging case with only 174 spots, JRSO substantially improved treatment quality by decreasing the objective function value from 1.71 to 0.339, reducing the maximum dose from 140.9% to 121.2%, enhancing the CI from 0.64 to 0.80, increasing FLASH coverage from 80.3% to 86.9%, and lowering the ROI effective dose from 6.23 to 6.02. These findings are further illustrated in Figure 3, which includes dose distribution maps and DVH for both JRSO and the conventional method. With the advantages of JRSO becoming more pronounced as the number of spots decreases. This is consistent with the conclusions drawn from the previous examples.

*3.3. HN*

The comparison between the conventional method and the proposed JRSO for the HN

cancer is presented in Table 4. For the case with 174 spots, JRSO reduced the objective function value from 0.254 to 0.242, decreased the maximum dose from 105.6% to 105.2%, and improved the CI from 0.87 to 0.88. In the scenario with 109 spots, JRSO exhibited even greater improvements, lowering the objective function value from 0.455 to 0.261, reducing the maximum dose from 117.6% to 107.1%, and enhancing the CI from 0.74 to 0.87. The most challenging case, with only 77 spots, further highlighted the effectiveness of JRSO. In this setting, JRSO significantly reduced the objective function value from 10.1 to 0.371, lowered the maximum dose from 134.0% to 113.1%, and substantially improved the CI from 0.41 to 0.70. Regarding FLASH coverage and ROI effective dose, both methods achieved high FLASH coverage, with the conventional method yielding slightly higher values by 1.3%, 2.9% and 6.2% for the three case respectively. However, JRSO consistently produced a superior ROI effective dose due to the higher-quality plans it generated. Specifically, JRSO reduced the ROI effective dose from 6.27 to 6.16 in the 174-spot case, from 6.50 to 6.10 in the 109-spot case, and from 7.55 to 6.27 in the 77-spot case. The DVH and dose plots in Figure 4 further validate the result.

**Table 2.** Prostate case. The dosimetric quantities from left to right are the optimization objective value, max target dose, conformal index (CI), FLASH coverage in percentage to the volume of ROI, and ROI effective dose.

|  |  | Objective | $D_{max}$ | CI | Coverage | Effective dose |
|---|---|---|---|---|---|---|
| 154 Spots | Conv | 0.143 | 120.6 | 0.79 | 93.4 | 6.33 |
|  | JRSO | 0.131 | 109.6 | 0.82 | 93.5 | 6.30 |
| 122 Spots | Conv | 0.162 | 132.8 | 0.80 | 94.2 | 6.40 |
|  | JRSO | 0.133 | 109.7 | 0.83 | 94.6 | 6.38 |
| 98 Spots | Conv | 0.259 | 121.8 | 0.68 | 91.4 | 6.41 |
|  | JRSO | 0.231 | 117.3 | 0.73 | 91.6 | 6.32 |

**Table 3.** Lung case. The dosimetric quantities from left to right are the optimization objective value, max target dose, conformal index (CI), FLASH coverage in percentage to the volume of ROI, and ROI effective dose.

|  |  | Objective | $D_{max}$ | CI | Coverage | Effective dose |
|---|---|---|---|---|---|---|
| 436 Spots | Conv | 0.249 | 116.0 | 0.86 | 93.7 | 6.33 |
|  | JRSO | 0.218 | 110.1 | 0.89 | 93.1 | 6.19 |
| 222 Spots | Conv | 0.264 | 114.7 | 0.83 | 91.0 | 6.22 |
|  | JRSO | 0.227 | 111.4 | 0.88 | 92.8 | 6.18 |
| 174 Spots | Conv | 1.71 | 140.9 | 0.64 | 80.3 | 6.23 |
|  | JRSO | 0.339 | 121.2 | 0.80 | 86.9 | 6.02 |

**Table 4.** HN case. The dosimetric quantities from left to right are the optimization objective value, max target dose, conformal index (CI), FLASH coverage in percentage to the volume of ROI, and ROI effective dose.

|  |  | Objective | $D_{max}$ | CI | Coverage | Effective dose |
|---|---|---|---|---|---|---|
| 174 Spots | Conv | 0.254 | 105.6 | 0.87 | 95.8 | 6.27 |
|  | JRSO | 0.242 | 105.2 | 0.88 | 94.5 | 6.16 |
| 109 Spots | Conv | 0.455 | 117.6 | 0.74 | 96.1 | 6.50 |
|  | JRSO | 0.261 | 107.1 | 0.87 | 93.2 | 6.10 |
| 77 Spots | Conv | 10.1 | 134.0 | 0.41 | 98.7 | 7.55 |
|  | JRSO | 0.371 | 113.1 | 0.70 | 92.5 | 6.27 |

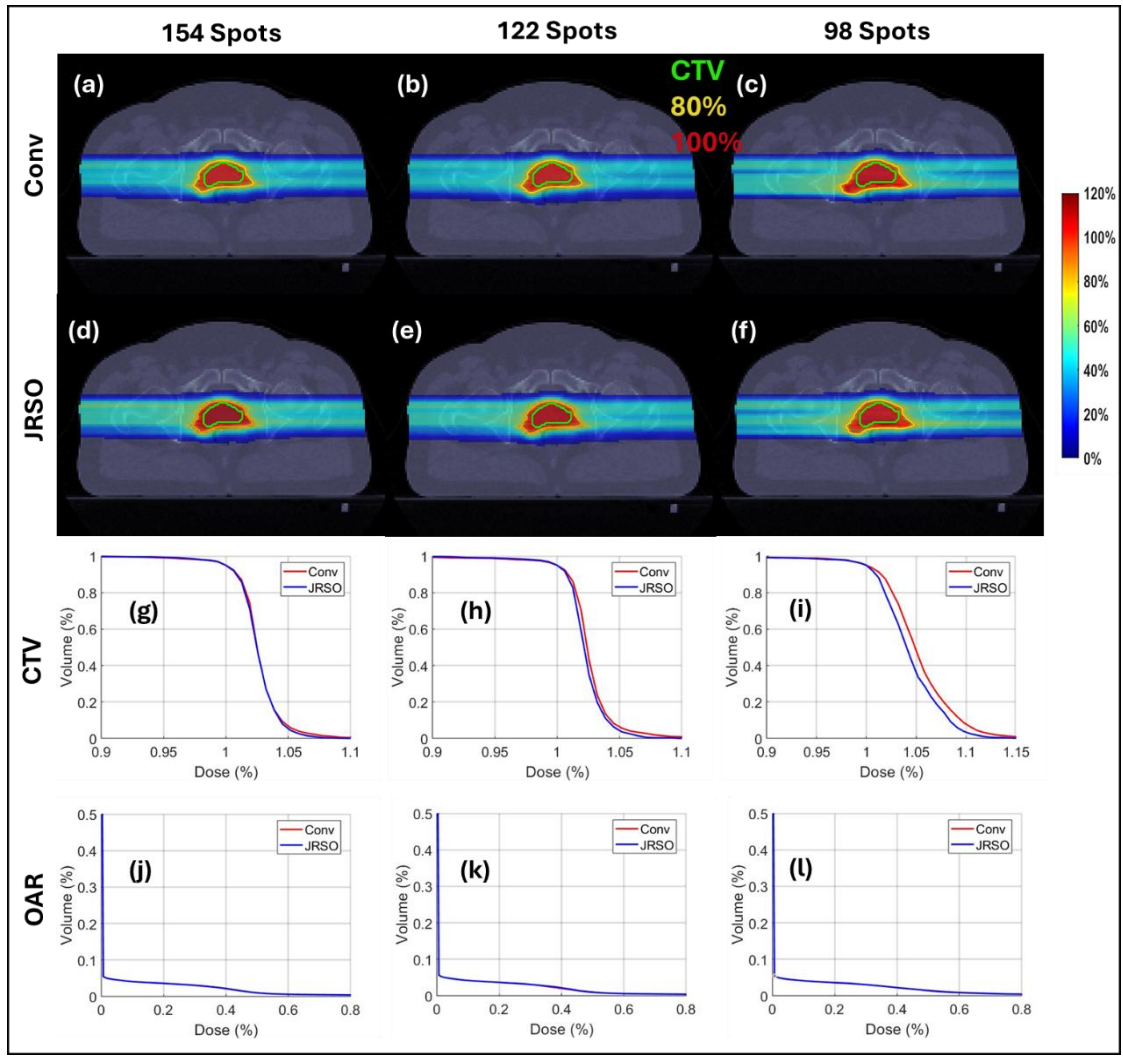

**Figure 2.** Prostate case. (a)-(c) and (d)-(f) are the dose plot of conventional method and JRSO method respectively. (g)-(i) and (j)-(l) are the DVH plots of CTV and OAR respectively. The dose plot window is set to [0%, 120%]. Isodose lines of 80%, 100% and the CTV are highlighted in the dose plot.

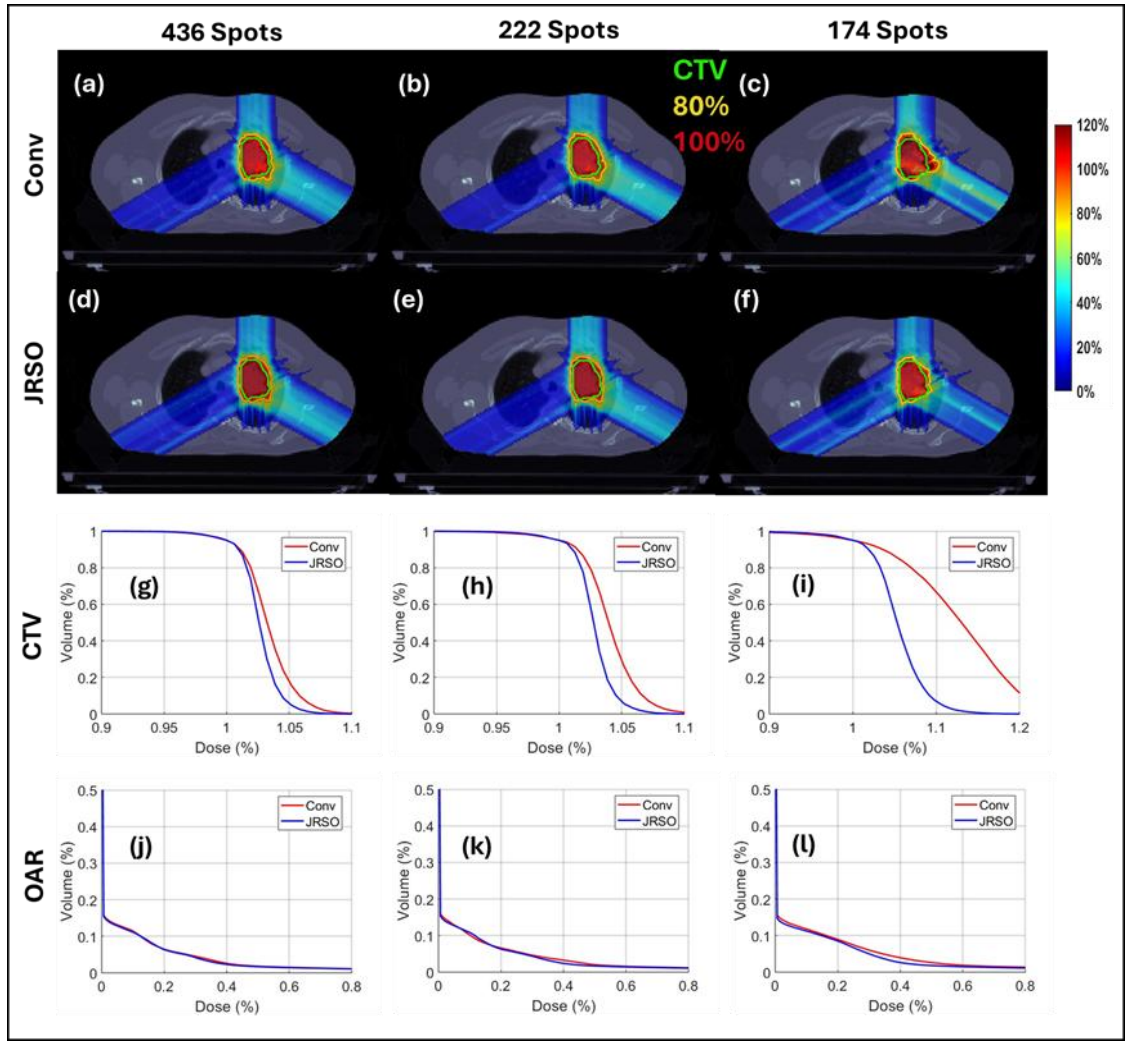

**Figure 3.** Lung case. (a)-(c) and (d)-(f) are the dose plot of conventional method and JRSO method respectively. (g)-(i) and (j)-(l) are the DVH plots of CTV and OAR respectively. The dose plot window is set to [0%, 120%]. Isodose lines of 80%, 100% and the CTV are highlighted in the dose plot.

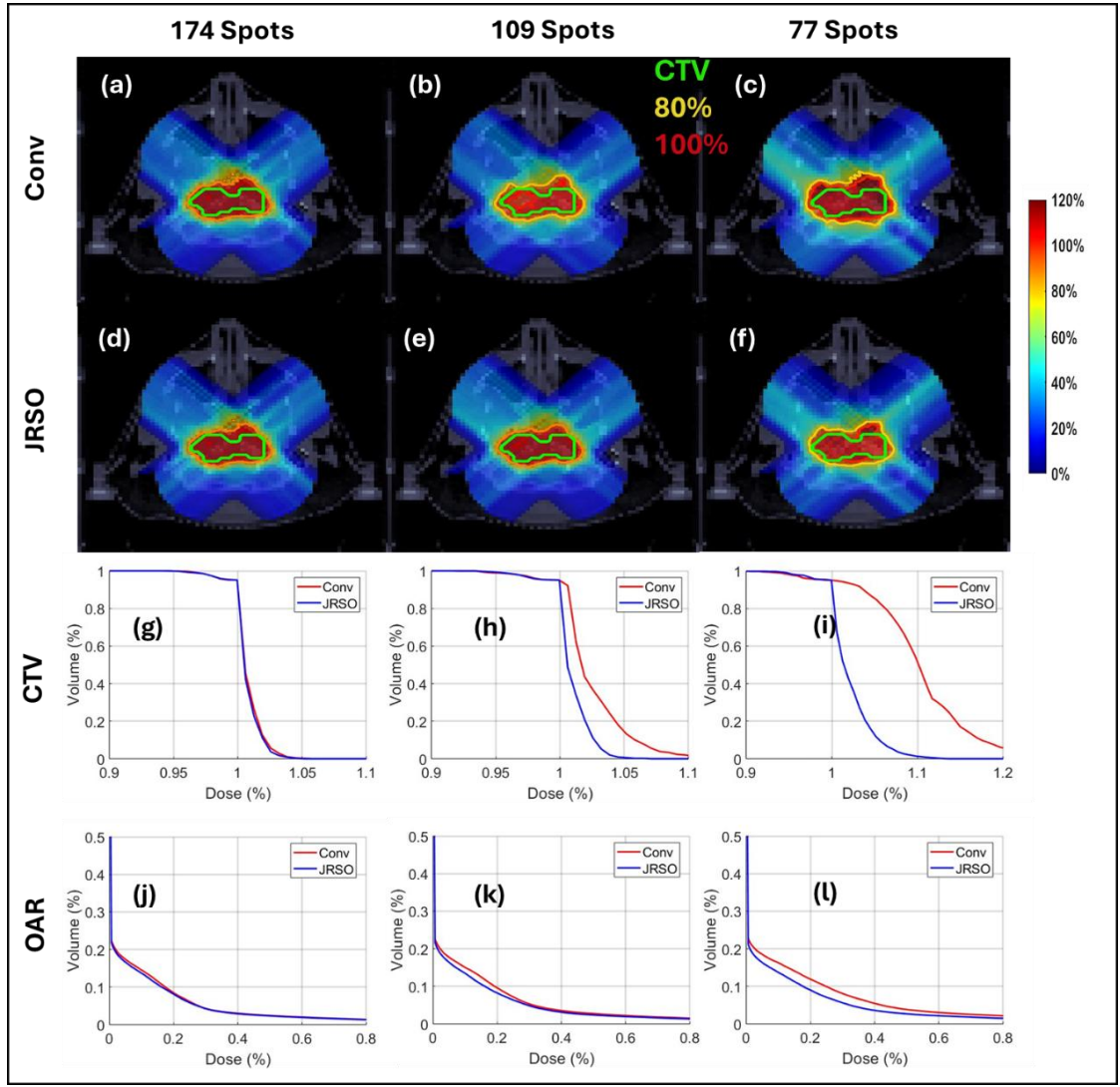

**Figure 4.** HN case. (a)-(c) and (d)-(f) are the dose plot of conventional method and JRSO method respectively. (g)-(i) and (j)-(l) are the DVH plots of CTV and OAR respectively. The dose plot window is set to [0%, 120%]. Isodose lines of 80%, 100% and the CTV are highlighted in the dose plot.

## 4. Discussion

This study presents JRSO, an optimization framework that simultaneously optimizes both the PSRM design and spot weights. As demonstrated in previous sections, this approach enhances plan quality and robustness compared to the conventional method.

While the main text has primarily focused on the overall performance of JRSO, one important aspect that has not yet been addressed is the impact of the proposed initialization strategy. Specifically, the framework integrates Algorithm 4 (see Appendix) as a preliminary step before initiating the alternating optimization process. Our preliminary numerical experiments reveal that, in the absence of proper initialization, the alternating optimization may converge very slowly or even diverge. This is due to the non-convexity of the problem. If the fixed parameter is poorly initialized, the optimization of the other parameter may steer the solution away from the true optimum. And such errors can accumulate across iterations, leading to instability or divergence.

By incorporating the initialization step via Algorithm 4, the optimization process is significantly stabilized. Although Algorithm 4 does not necessarily yield the globally optimal solution, it provides a solution that is sufficiently close to the optimum. This proximity ensures that the alternating optimization process starts from a favorable position, ultimately leading to reliable and efficient convergence toward the optimal solution.

Despite its demonstrated effectiveness, this study has certain limitations: Firstly, the current implementation of JRSO optimizes dose distribution but does not explicitly incorporate dose rate optimization. This results in a smaller FLASH coverage in some cases. For example, in the lung cancer case with 436 spots, the FLASH coverage achieved by the conventional method is 93.7%, while JRSO achieves 93.1%. Similarly, in the HN case, the conventional method maintains higher FLASH coverage across all three scenarios. While JRSO compensates for this with superior ROI effective dose, future work could integrate dose rate optimization, potentially leveraging methodologies such as those proposed in [30].

Secondly, the current numerical experiments focus primarily on CTV planning, however, the JRSO framework can be readily extended to include OAR considerations, which will be explored in future research.

Thirdly, while the numerical simulations demonstrate the robustness and efficacy of JRSO, further validation through experimental and clinical studies is essential to assess its real-world applicability. Future research will focus on implementing JRSO in experimental settings and evaluating its performance under realistic treatment conditions.

## 5. Conclusion

This study proposes a novel alternating optimization method JRSO for pFLASH radiotherapy. The new method may extend the applicability of PSRM to more complex clinical scenarios, particularly those involving misalignments between beams and pins. Numerical results demonstrate the robustness and efficiency of the new method.

**Appendix**

*1. ADMM*

In this section, we give a detailed derivation of Algorithm 2,3 using ADMM. Recall equation (6)

$$\min_{x} G_{a^{k-1}}(x) = \frac{1}{2} \left\| A^k \left( \Omega(x, a^{k-1}) \right) x - b \left( \Omega(x, a^{k-1}) \right) \right\|^2, \quad (A.1)$$

$$\text{s.t. } x_i \in \{0\} \cup [g, +\infty).$$

Introducing the auxiliary variable $\mu$ and let $I_{C(g)}(x)$ be the indicator function for set $C(g) = \{0\} \cup [g, +\infty)$,

$$I_{C(g)}(x) = \begin{cases} 0, & \text{if } x \in C(g) \\ +\infty, & \text{otherwise} \end{cases}$$

(A.1) is equivalent to

$$\min_{x,\mu} \frac{1}{2} \left\| A^k \left( \Omega(x^{m-1}, a^{k-1}) \right) x - b \left( \Omega(x^{m-1}, a^{k-1}) \right) \right\|^2 + I_{C(g)}(\mu), \quad (A.2)$$

$$\text{s.t. } x - \mu = 0.$$

To deal with the convex optimization with equality constraint, we consider the augmented Lagrangian function of (3):

$$L(x, \mu, \lambda) = \frac{1}{2} \left\| A^k \left( \Omega(x^{m-1}, a^{k-1}) \right) x - b \left( \Omega(x^{m-1}, a^{k-1}) \right) \right\|^2$$

$$+ I_{C(g)}(\mu) + \frac{\rho}{2} \left\| x - \mu + \frac{\lambda}{\rho} \right\|^2. \quad (A.3)$$

Let $\lambda = \frac{\lambda}{\rho}$, the ADMM takes the form of:

$$x^m = \underset{x}{\operatorname{argmin}} \frac{1}{2} \left\| A^k \left( \Omega(x^{m-1}, a^{k-1}) \right) x - b \left( \Omega(x^{m-1}, a^{k-1}) \right) \right\|^2 + \frac{\rho}{2} \| x - \mu^{m-1} + \lambda^{m-1} \|^2, \quad (A.4)$$

$$\mu^m = \underset{\mu}{\operatorname{argmin}} I_{C(g)}(\mu) + \frac{\rho}{2} \| x^m - \mu + \lambda^{m-1} \|^2, \quad (A.5)$$

$$\lambda^m = \lambda^{m-1} + x^m - \mu^m. \quad (A.6)$$

solved analytically. Applying the first-order optimality condition, $x^m$ satisfies

$$A^k \left( \Omega(x^{m-1}, a^{k-1}) \right)^T \left( A^k \left( \Omega(x^{m-1}, a^{k-1}) \right) x^m - b \left( \Omega(x^{m-1}, a^{k-1}) \right) \right)$$

$$+\rho(x^m - \mu^{m-1} + \lambda^{m-1}) = 0,$$

which is equivalent to

$$\left(A^k\left(\Omega(x^{m-1}, a^{k-1})\right) + \rho I\right)x^m$$

$$= A^k\left(\Omega(x^{m-1}, a^{k-1})\right)^T b\left(\Omega(x^{m-1}, a^{k-1})\right) + \rho(\mu^{m-1} - \lambda^{m-1}). \tag{A.7}$$

For the second subproblem (6), it is the same as

$$\mu^m = \underset{\mu \in C(g)}{\operatorname{argmin}} \|x^m - \mu + \lambda^{m-1}\|^2,$$

According to the definition of $C(g)$,

$$\mu_i^m = \begin{cases} \max(x_i^m + \mu_i^{m-1}, g), & \text{if } x_i^m + \mu_i^{m-1} \geq \frac{g}{2} \\ 0, & \text{otherwise}. \end{cases} \tag{A.8}$$

Where $i$ denotes the $i$-th component. And Algorithm 2 consists of equation (A.6), (A.7) and (A.8).

*2. Initialization strategy*

As the problem is relatively highly nonconvex, the alternative optimization process mentioned before may be sensitive to initial condition. Inspired by the previous works on FLASH [18-20] in which they generate a reference IMPT plan first and then convert the energy levels into slabs of range modulator, we propose the following initialization strategy.

$$\min_z G(z) = \sum_{s=1}^{S} \omega_s \left\|D_{\Omega_s(z)} z - d_{\Omega_s(z)}\right\|^2, \tag{A.9}$$

$$\text{s.t. } F^\dagger Cz \in \{0\} \cup [g, +\infty).$$

Here $F^\dagger$ is the pseudoinverse of $F$, $z$ is the spot weight coming out of the PSRM. By the definition of pseudoinverse, $F^\dagger Cz$ is the projection of $Cz$ onto the range space of $F$, and $F^\dagger Cz$ is the optimal approximation of $x$ in $L_2$ sense. As a result, $F^\dagger Cz$ should satisfy the MMU constraint. We rewrite the formulation above as below:

$$\min_z G_z(z) = \frac{1}{2}\|A(\Omega(z))z - b(\Omega(z))\|^2, \tag{A.10}$$

$$\text{s.t. } F^\dagger Cz \in \{0\} \cup [g, +\infty).$$

with $\Omega(z)$ being the active index based on $z$.

The optimization of (11) can be written as the algorithm below:

---
**Algorithm 4**: ADMM for optimizing $z$
---
**Input**: $z^0, \mu^0, \lambda^0$, MMU threshold $g$, proper $\rho > 0$.
  **For** $m = 1,2,\ldots,M$
  $$z^m = \left(A^{m-1}(\Omega(z^{m-1}))^T A^{m-1}(\Omega(z^{m-1})) + \rho I\right)^{-1}$$
  $$\left(A^{m-1}(\Omega(z^{m-1}))^T b(\Omega(z^{m-1})) + \rho(\mu^{m-1} - \lambda^{m-1})\right).$$
  $$\mu^m = \begin{cases} \left(I_H F(max(z^m + \lambda^{m-1}, g))\right) \circ (z^m \oslash (\hat{C}z^m)), & \left(F^\dagger C\right)(z^m + \lambda^{m-1}) \geq \frac{g}{2}, \\ 0, & \text{otherwise.} \end{cases}$$
  $$\lambda^m = \lambda^{m-1} + z^m - \mu^m.$$
  **End**
**Output** $z^M$.

---

After the solution of (A.10) is obtained, $x$ and $a$ are solved accordingly. $x^0$ is computed by:

$$x^0 = \left(F^\dagger C\right) z^M.$$

To calculate $a^0$ according to $z^M$, the matrix $\hat{C}$ is introduced:

$$\hat{C} = \begin{bmatrix} \mathbb{1}_{H\times H} & \mathbb{0}_{H\times H} & \cdots & \mathbb{0}_{H\times H} \\ \mathbb{0}_{H\times H} & \mathbb{1}_{H\times H} & & \mathbb{0}_{H\times H} \\ \vdots & & \ddots & \vdots \\ \mathbb{0}_{H\times H} & \mathbb{0}_{H\times H} & \cdots & \mathbb{1}_{H\times H} \end{bmatrix}_{N\times N}.$$

And $a$ is calculated by

$$a^0 = z^M \oslash (\hat{C}z^M),$$

where $\oslash$ is the elements-wise division.

This will provide a good initialization for the following alternating optimization process. It is important to note, however, that the subsequent alternating optimization step is crucial. This is because, in one-to-many correspondence scenarios, the pseudoinverse of $F$ does not always yield an $x$ that satisfies the MMU constraint. Alternating optimization is needed to ensure that the final solution adheres to all clinical and technical requirements. Note that in the numerical experiments of section 3, the conventional method used for comparison corresponds to Algorithm 4. However, instead of only providing a rough initialization, the algorithm is executed until full convergence is achieved, ensuring a precise and optimized result for comparison.